\documentclass[prl,twocolumn,showpacs,showkeys,preprintnumbers,aps,amssymb]{revtex4}

\usepackage{amsmath,amssymb}
\usepackage{graphicx}
\usepackage{dcolumn}
\usepackage{bm}

\newcommand{\bfm}[1]{\mbox{\boldmath$#1$}}

\def\ben{\begin{equation}}
\def\een{\end{equation}}
\def\bey{\begin{eqnarray}}
\def\eey{\end{eqnarray}}
\def\ba{\begin{array}}
\def\ea{\end{array}}
\def\benmrt{\begin{enumerate}}
\def\eenmrt{\end{enumerate}}

\def\psla{p{\raise1pt\hbox{$\!\!/$}}}
\def\dsla{\partial{\raise1pt\hbox{$\!\!\!/$}}}
\def\Dsla{D{\raise1pt\hbox{$\!\!\!/$}}}
\def\xsla{x{\raise1pt\hbox{$\!\!\!/$}}}

\def\jmu5{j_{\mu 5}^{(i)}(0)}
\def\jnu5{j_{\nu 5}^{(i)}(0)}

\def\qq0v{\langle0\!\mid\!{\bar q}q\!\mid\! 0\rangle}

\def\qc0f{\langle0\!\mid\!{\bar q}q\!\mid\!0\rangle_{F}}

\def\qsq0f{\langle0\!\mid\!{\bar q}\sigma_{\mu\nu}q\!\mid\!0\rangle_{F}}

\def\qgdq0f{\langle0\!\mid\!{\bar q}{\cal 
S}\gamma_{\mu}D_{\nu}q\!\mid\!0\rangle_{F}}

\def\qddq0f{\langle0\!\mid\!{\bar q}{\cal
S}D_{\mu}D_{\nu}q\!\mid\!0\rangle_{F}}

\def\3mmtm{|{\bf q}|^2}

\def\eq#1{Eq.(\ref{#1})}
\def\eqs#1#2{Eqs.(\ref{#1}) and (\ref{#2})}

\def\Ref#1{[\ref{#1}]}
\def\Refs#1#2{[\ref{#1},\ref{#2}]}

\def\p0{p_0}

\def\gam3{\mbox{\boldmath{$\gamma$}}}
\def\e0{E_{0}(s_{0},s)}
\def\e1{E_{1}(s_{0},s)}
\def\e2{E_{2}(s_{0},s)}

\def\aplt{\kern0.3333em \raise 0.2ex \hbox{$<$}%
\kern-0.8em \lower0.8ex \hbox{$\sim$}%
\kern0.3333em}
\def\aplg{\kern0.3333em \raise 0.2ex \hbox{$>$}%
\kern-0.8em \lower0.8ex \hbox{$\sim$}%
\kern0.3333em}

\def\ln{{\rm ln}}

\begin{document}

\preprint{TH-881}

\title{Effect of pion thermal width on the sigma spectrum}

\author{Yoshimasa HIDAKA}
\email{hidaka@post.kek.jp}
\author{Osamu MORIMATSU}
\email{osamu.morimatsu@kek.jp}
\author{Tetsuo NISHIKAWA}
\email{nishi@post.kek.jp} 
\author{Munehisa OHTANI}
\email{ohtani@post.kek.jp} 
\affiliation{%
Institute of Particle and Nuclear Studies, 
High Energy Accelerator Research Organization, 1-1, Ooho, 
Tsukuba, Ibaraki, 305-0801, Japan
}

\date{April 24, 2003}

\begin{abstract}
We study the effect of thermal width of $\pi$ on the spectral function of $\sigma$
applying a resummation technique called optimized perturbation theory at finite temperature
($T$) to ${\cal O}(4)$ linear sigma model.
In order to take into account finite thermal width of $\pi$,
we replace the internal pion mass in the self-energy of $\sigma$ with that of complex pole found in a previous paper.
The obtained spectral function for $T\aplg 100\,{\rm MeV}$ turns out to possess two broad peaks.
Although a sharp peak at $\sigma\rightarrow\pi\pi$ threshold was observed in the one-loop calculation without pion thermal width, the peak is proved to be smeared out.
We also search for the poles of the $\sigma$ propagator
and analyze the behavior of the spectral function with these poles.

\end{abstract}
\maketitle

\newpage

It is now believed that the chiral symmetry is restored at finite temperature.
As temperature increases to the critical temperature,
$\sigma$ is softened while light $\pi$ becomes heavy,
because their masses should degenerate after the symmetry restoration.
At certain temperature, therefore, the mass of $\sigma$ coincides with 
twice that of $\pi$. Accordingly, the spectrum of $\sigma$
is expected to be enhanced near the threshold of $\sigma\rightarrow\pi\pi$,
since the phase space available for the decay is squeezed to zero.

Chiku and Hatsuda \cite{chiku} showed that the threshold enhancement in the $\sigma$ channel
is observed as expected.
They also calculated the spectral function of $\pi$ and found that $\pi$ has finite width 
due to the scattering with thermal pions in the heat bath:
$\pi+\pi^{\rm thermal}\rightarrow \sigma$.
The width of $\pi$ is $\sim 50\,{\rm MeV}$ at the temperature where the threshold in the $\sigma$ spectrum is most strongly enhanced.
In their analysis of the spectrum of $\sigma$, however,
the effect of the pion thermal width is not included,
since they calculated the self-energy only up to one-loop order.
The purpose of this paper is to show that the threshold enhancement is smeared out by taking into account the effect of pion thermal width.

Our strategy is as follows.
We utilize the one-loop self-energy for $\sigma$,
but we replace the masses of internal pions with complex ones.
This complex mass was obtained from the location of the pole of the pion propagator
in a previous work \cite{hidaka}.
Using the self-energy with the complex pion mass, 
we study the spectral function of $\sigma$.
The poles of the $\sigma$ propagator are also searched for
to analyze the behavior of the spectral function.

Along the lines mentioned above, here we employ
the ${\cal O}(4)$ linear sigma model:
\ben
  {\cal L}= \frac{1}{2}(\partial\phi_{\alpha})^2-\frac{1}{2}\mu^2 
  \phi_{\alpha}^{2}-
             \frac{\lambda}{4!}(\phi_{\alpha}^{2})^2 +h\phi_0 \; ,
\een
with $\phi_\alpha=(\phi_0,\bfm{\pi})$. 
As the chiral symmetry is taken to be dynamically broken at low temperature ($T$),
the quadratic term in the Lagrangian has negative sign, $\mu^2 < 0$.
Owing to this choice of the sign and the explicit breaking term 
of $h\phi_0$, the field $\phi_0$ has a non-vanishing 
expectation value $\xi$.
In this prospect, we decompose beforehand the field operator $\phi_0$ 
into the classical condensate and the quantum fluctuation as
$\phi_0 = \xi + \sigma\;$.  

Integrating out the quadratic of the fluctuations $\sigma$ and $\pi$
around the condensate, we obtain the one-loop effective potential
 $V^{\rm eff}(\xi)$. 
As a minimum of the effective potential, the condensate $\xi(T)$ 
is determined and this leads us to the gap equation,
\begin{align}
  0=\frac{\partial V^{\rm eff}}{\partial \xi}&=-h+\mu^2\xi
   +\frac{\lambda}{3!}\xi^3 
   +\frac{\lambda \xi}{2}(I^{(1)}_\sigma +I^{(1)}_\pi) \; .
\label{eq:gap}
\end{align}
The model parameters are fixed
so as to reproduce the experimental values of the pion mass ($m_{\pi}$), the pion decay constant ($f_\pi$)
and the peak energy of the spectral function 
$\rho_\sigma$ at $T=0$ following Ref.\cite{chiku}:
$\mu^2=-(283\,{\rm MeV})^2,\ \lambda=73.0, \ h=(123\,{\rm MeV})^3$.
The last term in \eq{eq:gap} is contribution
from the tadpole diagrams in the modified minimal subtraction scheme
($\overline{\rm MS}$):
\begin{equation}
  I^{(1)}_{\phi} 
       \stackrel{\overline{\rm MS}}{=} -\frac{m_{0\phi}^2}{16\pi^2}(
        1-\ln\frac{m_{0\phi}^2}{\kappa^2}) 
 +\int_0^\infty\frac{{\rm d}p\,p^2}{2\pi^2}
 \frac{n(\omega_\phi)}{\omega_\phi}\;,
  \label{eq:I1}
\end{equation}
with $\omega_\phi=\sqrt{p^2+m_{0\phi}^2}$ and
$n(\omega)=({\rm e}^{\omega/T}-1)^{-1}$.
$\kappa$ is the renormalization point.
$m_{0\sigma}$ and $m_{0\pi}$ are the tree-level masses of 
the fluctuations defined by
\[ m_{0\sigma}^2=m^2+\lambda \frac{\xi^2}{2} \; ,
 \ \ m_{0\pi}^2=m^2+\lambda \frac{\xi^2}{6} \; ,
\] 
where $m(T)$ is an optimal mass parameter
introduced in the framework of the optimized perturbation theory (OPT) \cite{chiku,stev}.

A physical concept of OPT is to impose 
thermal effects upon the optimal parameter(s) and to tune it
by an optimal condition \cite{stev}.
In our case, $m$ is determined so that the (thermal part of) pion-mass correction
vanishes at zero momentum \cite{chiku}.

Making use of the OPT, the threshold enhancement of the spectral function
of $\sigma$ was reported \cite{chiku}. 
In order to understand how this enhancement occurs and will be smeared 
by the thermal width of $\pi$, it is important to consider 
the scattering with thermal pions as mentioned previously.
 At the same time, it is also of much account
to clarify analyticity of the self-energy as well.
This is basically because
the spectral function, which is the imaginary part of 
the propagator, is related to the self-energy $\Pi_\sigma(k,T)$ 
as
\bey
  \label{eq:spec}
\rho_\sigma(k,T)&=& 2\ {\rm Im}\ D_\sigma(k,T)\cr
&=&
 -2 \ {\rm Im}\ (k^2-m_{0\sigma}^2-\Pi_\sigma(k,T))^{-1}\; ,
\eey
with $k=(\omega,\boldsymbol{k})$. 
At one-loop level, the self-energy is written as
\bey
  \Pi_\sigma(k,T) &=& -(m^2-\mu^2)+\frac{\lambda}{2}\left(I^{(1)}_\sigma
    +I^{(1)}_\pi\right)\cr
    &&-\frac{\lambda^2\xi^2}{2}\left(
 I^{(2)}_\sigma+\frac{1}{3}I^{(2)}_\pi\right) \; ,
  \label{eq:selfen}
\eey
where $I^{(2)}_\phi$ corresponds to a bubble diagram shown in Fig.1,
\begin{align}
I^{(2)}_{\phi} &\stackrel{\overline{\rm MS}}{=}
   \frac{1}{16\pi^2}\left(2-\ln\frac{m_{0\phi}^2}{\kappa^2}
    +c\,\ln \frac{c-1}{c+1}   \right) \nonumber
   \\ & \hspace{3em} 
 -\frac{1}{\pi^2}\int_0^\infty\frac{{\rm d}p \,p^2}{\omega_\phi}
   \frac{n(\omega_\phi)}{k^2-4\omega_\phi^2}\; ,
  \label{eq:I2}
\end{align}
with $c=\sqrt{1-4m_{0\phi}^2/k^2}$ .

Now, we are ready to consider effects of  
pion thermal width on the spectrum of $\sigma$.
Although the width can be estimated directly
from relevant scattering processes,
we adopt more simple prescription --- to use a complex pion pole
as internal pion mass in $\Pi_{\sigma}$.
The thermal width of $\pi$ 
is represented in the imaginary part of 
the complex pole of the pion propagator:
\begin{align*}
-D_\pi^{-1} & = \left. p^2-m_{0\pi}^2
-\Pi_\pi(p^2, {\bfm{p}}\; ; T) \right|_{p^2=(m_\pi^{\rm pole})^2}=0 
\; , \\  
& \ \ m_\pi^{\rm pole}=m_\pi^{*}(T)-i \frac{\varGamma_\pi(T)}{2} \; ,
\end{align*}
where $\Pi_{\pi}$ is the pion self-energy. 

\begin{figure}
\begin{center}
\includegraphics[keepaspectratio,height=3cm]{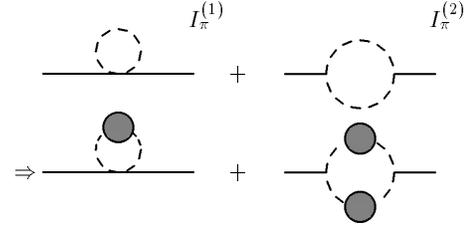}
\caption{A Prescription to take into account the thermal width of $\pi$ in the decay $\sigma\rightarrow\pi\pi$. The upper diagrams represent the self-energy of $\sigma$, \eqs{eq:I1}{eq:I2}, with the internal pion mass, $m_{0\pi}$, and the lower diagrams with $m_{\pi}^{\rm pole}$.}
\label{diagram}
\end{center}
\end{figure}

As a matter of course, the pole has dependence 
of $|\boldsymbol{p}|$ naturally \cite{hidaka},
but contributions from $|\boldsymbol{p}|\gtrsim T$ 
are less effective to the loop integration in the thermal part of $\Pi_\sigma$.
Therefore, as a simple recipe to incorporate 
the thermal width,
we replace $m_{0\pi}$ in \eq{eq:selfen} with $m_\pi^{\rm pole}$ at $\bfm{p}={\bf 0}$ 
in Ref.\cite{hidaka} as depicted in Fig.\ref{diagram} 
and substitute it into \eq{eq:spec} to obtain $\rho_\sigma$.

\begin{figure}
\begin{center}
\includegraphics[width=8.5cm]{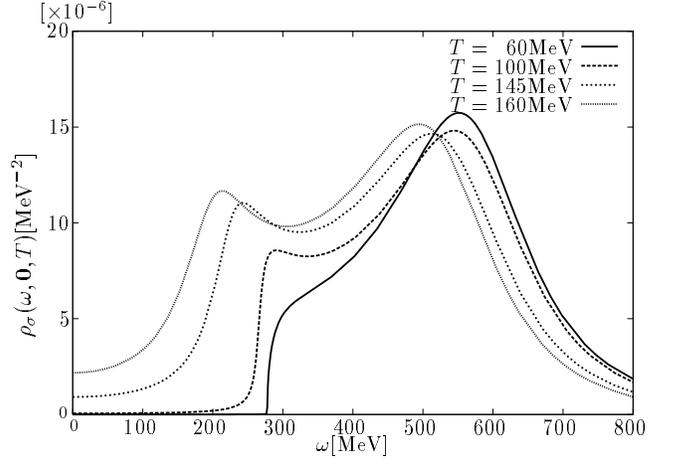}
\caption{Spectral function of $\sigma$, \eq{eq:spec}, with the thermal width of $\pi$.}
\label{spectral}
\end{center}
\end{figure}

\begin{figure}
\begin{center}
\includegraphics[width=8.5cm]{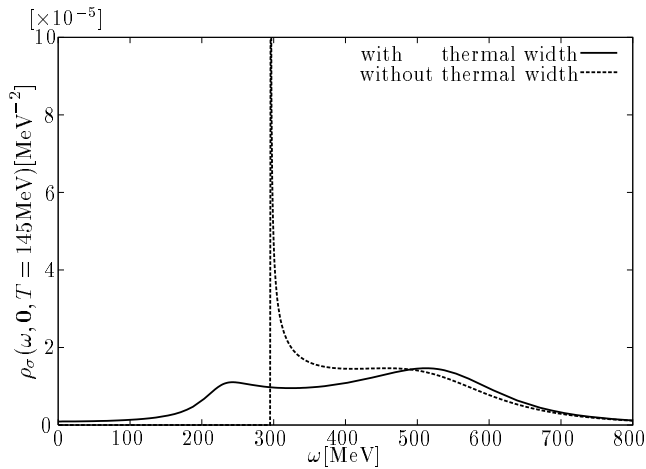}
\caption{Spectral function of $\sigma$ with and without the thermal width of $\pi$ at $T=145\,{\rm MeV}$.}
\label{hikaku}
\end{center}
\end{figure}

We show in Fig.\ref{spectral} the obtained spectral function, \eq{eq:spec}, at ${\bfm k}={\bf 0}$ for several values of $T$.
For comparison, we also show the spectral function at $T=145\,{\rm MeV}$
with $m_{0\pi}$ and that with $m_\pi^{\rm pole}$ in Fig.\ref{hikaku}.
At low $T$ the spectrum consists of a broad bump 
around $\omega=550\,{\rm MeV}$.
As $T$ increases, another strength grows
at the left shoulder of the bump,
while the spectrum for $\omega\aplg 400\,{\rm MeV}$ does not vary much.
In contrast to this case, 
if $m_{0\pi}$ is used as the internal pion mass,
the spectrum near the threshold of $\sigma\rightarrow\pi\pi$ is enhanced as $T$ is increased. 
At $T=145\,{\rm MeV}$, where the mass of $\sigma$ coincides with $2m_{0\pi}$,
the spectrum near the threshold
is most strongly enhanced as shown in Fig.\ref{hikaku} (dotted curve).
When $m_{\pi}^{\rm pole}$ is used, however, 
due to the thermal width of $\pi$ in the decay $\sigma\rightarrow\pi\pi$,
the sharp threshold is smeared to be a bump at $\omega=200\,{\rm MeV}\sim 300\,{\rm MeV}$ as the solid curve in Fig.\ref{hikaku}.

In general, the behavior of spectral function is governed 
by poles of the propagator.
We thus next search for the poles of propagator on the complex $\omega$ plane.
As can be clearly seen from \eq{eq:spec} the analytic structure of the propagator is determined by that of the self-energy.
Then, from \eq{eq:selfen} and \eq{eq:I2} we see that the self-energy has a branch point at $\omega=2m_\pi^{\rm pole}$ around which the Riemann sheet is two-fold.
The self-energy has another branch point due to $\sigma$ loop which, however, is irrelevant for the present discussion of the self-energy around the threshold of $\sigma\rightarrow\pi\pi$.
The branch cut from $\omega=2m_\pi^{\rm pole}$ can be arbitrarily chosen at finite temperature, because it has no physical meaning.
For definiteness we choose the cut from the branch point straight to the right, parallel to the real axis and refer to the sheet including the physical real energy as the first Riemann sheet and the other one as the second Riemann sheet.
We show in Fig.{\ref{poleNZwidth} the result of searching the locations of the poles on the Riemann sheets.
\begin{figure}
\begin{center}
\includegraphics[width=8.5cm]{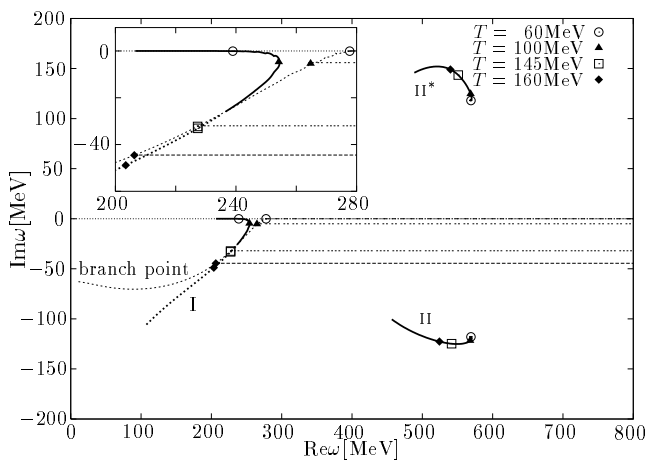}
\caption{Location of the poles of the $\sigma$ propagator with the thermal width of $\pi$,
$m_{\sigma}^{\rm pole}$.
The branch point is shown by dotted line 
and the branch cut are also indicated.}
\label{poleNZwidth}
\end{center}
\end{figure}

We found three relevant poles.
Two of them, (II) and (II*), are on the second Riemann sheet
and off the real axis. Their positions are approximately symmetric 
with respect to the real axis with each other.
For the spectral function of $\sigma$, on one hand,
(II) and (II*) show up as a bump around $\omega=550\,{\rm MeV}$ and
the small movement of these poles explains the insensitivity 
to $T$ of the spectrum for $\omega \gtrsim 400\,{\rm MeV}$.
On the other hand, the structure of the spectrum 
for low $\omega$ is mainly determined by
the remaining one pole, (I).
For preparation to consider this pole (I), 
let us first recall 
the movement of 
the corresponding pole in case of no thermal width 
of $\pi$ \cite{hidaka}.

Without the pion width,
the counterpart of the pole (I)
exists on the second Riemann sheet apart from the branch point of
$2m_{0\pi}$ for $T \lesssim 60\,{\rm MeV}$.
Correspondingly, the $\sigma$ spectrum has no peak
at the $\pi\pi$ threshold for low $T$.
As $T$ increases, the pole moves along the real axis
from below toward the branch point $2m_{0\pi}$ and turns around 
at the point to appear on the first sheet at $T=145\, {\rm MeV}$.
In response to this movement, the $\sigma$ spectrum
is strongly enhanced for $T \sim 145 \,{\rm MeV}$.
In other words, the threshold enhancement 
originates in the fact that the $\sigma$ pole approaches 
the branch point on the real axis near the temperature concerned.

Keeping this in mind, 
we return to the discussion on movement of 
the pole (I) next. 
 As is the case with no pion width,
this pole also resides on the second Riemann sheet for low $T$
and comes closer to the branch point with increasing $T$ 
as shown in Fig.{\ref{poleNZwidth}.
At $T=138\,{\rm MeV}$, the pole crosses the branch cut and appears
on the first Riemann sheet from the second, likewise.
%
 The branch point, however,
is located at $2m_\pi^{\rm pole}$ apart from the real axis
for $T\gtrsim 100 \, {\rm MeV}$ due to the finite thermal width of $\pi$ 
(see Fig.{\ref{poleNZwidth}).
The complex branch point demands that 
the pole (I), whose counterpart caused the threshold enhancement
by moving toward the point, 
also acquires imaginary part near $T \sim 138\,{\rm MeV}$.
As a result, the $\sigma$ spectrum is smeared and
the remarkable contrast is brought about
for the spectrums as shown in Fig.\ref{hikaku}.

Let us argue on the correlation between the spectral function and poles 
of the propagator more quantitatively.
For this purpose,
we approximate the propagator of $\sigma$ with
the superposition of the three pole contributions in the following way:
\bey
&&\rho_\sigma \simeq -2{\rm Im} \sum_{\rm pole} \frac{Z^{\rm pole}}{q-q_{\rm pole}},\cr
&&
Z^{\rm pole} = \frac{1}{2q_{\rm pole}}
\left(  1-\frac{\partial\Pi_\sigma}{\partial k^2}
 (k^2=(m_\sigma^{\rm pole})^2)\right)^{-1},
\label{approx-rho}
\eey
where $m_\sigma^{\rm pole}$ stands for the pole of the $\sigma$ propagator.
$q$ and $q_{\rm pole}$ are defined by $q=(k^2-(2m_\pi^{\rm pole})^2)^{1/2}$
and $q_{\rm pole}=((m_{\sigma}^{\rm pole})^2-(2m_\pi^{\rm pole})^2)^{1/2}$, respectively.
It should be noted that we have taken not $k^2$
but $q=(k^2-(2m_\pi^{\rm pole})^2)^{1/2}$ as an appropriate variable.
The use of this variable enables us to
unfold two leaves of the Riemann sheet around the branch point \cite{newton};
We can reflect which Riemann sheet accommodates 
each pole on the approximated propagator.
\begin{figure}
\begin{center}
\includegraphics[width=8.5cm]{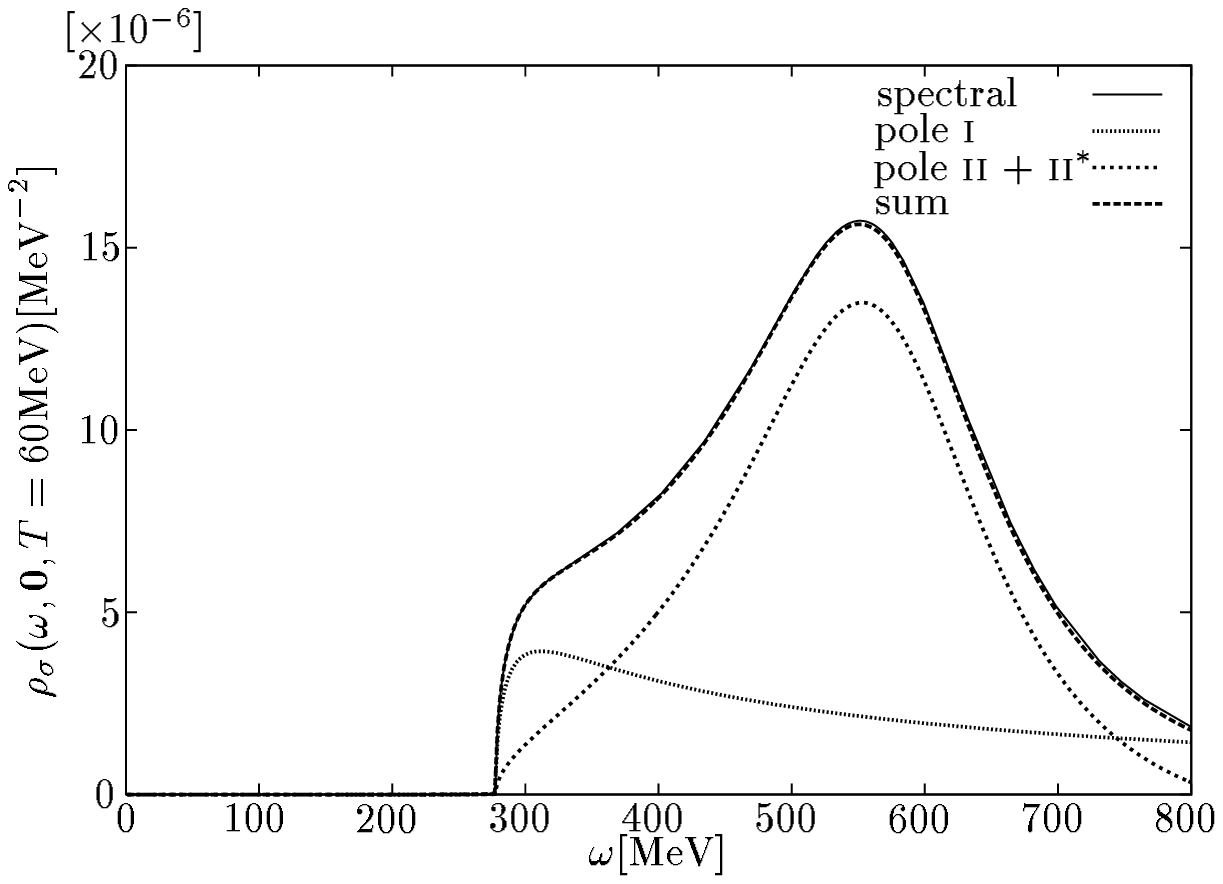}
\caption{The spectral function approximated by a superposition of the contributions from the three poles, \eq{approx-rho}, at $T=60\,{\rm MeV}$.
The spectral function, \eq{eq:spec}, is also shown.}
\label{polefit60}
\end{center}
\begin{center}
\includegraphics[width=8.5cm]{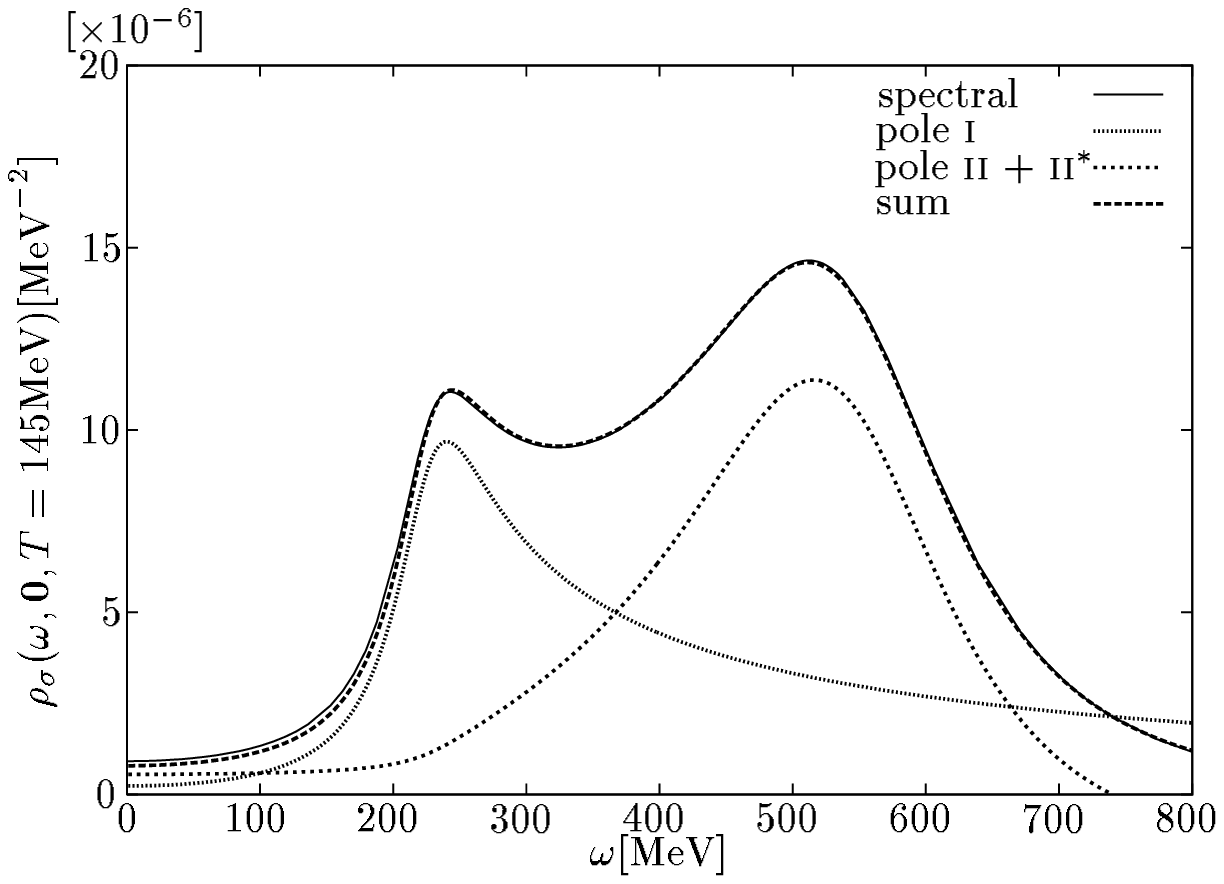}
\caption{The same as in Fig.\ref{polefit60} at $T=145\,{\rm MeV}$.}
\label{polefit145}
\end{center}
\end{figure}

We show the approximated spectral function, \eq{approx-rho}, 
for two cases, $T=60\,{\rm MeV}$ and $T=145\,{\rm MeV}$,
in Figs.\ref{polefit60} and \ref{polefit145}, respectively.
At $T=60\,{\rm MeV}$, pole (I) is located on the second Riemann sheet,
while it is on the first sheet at $T=145\,{\rm MeV}$.
We can see that poles (II) and (II*) provide the broad peak around $\omega=550\,{\rm MeV}$ and that pole (I) mainly determine 
the behavior of the left shoulder of the peak.
If we include only the contributions of (I) and (II), 
two broad peaks are still reproduced.
However, a structure which is not seen in the spectral function appears at low $\omega$ region.
The contribution of (II*) cancels out the spurious structure caused by (II).  
At $T=60\,{\rm MeV}$, the left shoulder of the peak is small,
although pole (I) seems to lie close to the real axis.
This is because the pole is on the second Riemann sheet 
and is practically far from the real axis.
In contrast, at $T=145\,{\rm MeV}$, pole (I) is on the first Riemann sheet
and is close to the real axis.
Accordingly, it provides a peak with more strength around $\omega=200\sim300\,{\rm MeV}$.
Thus, in order to reveal the relation 
between the poles and the behavior of the spectrum,
it is necessary to know both of the complex structure of the Riemann sheet 
and the positions of the poles on it.

In conclusion, the thermal width of $\pi$ smears out the sharp peak of the $\sigma$ spectrum at $\pi\pi$ threshold.
The structure of the spectral function
can be well understood from the three relevant poles of the propagators 
and the branch points of the self-energy.
Physically we can interpret the origin of the smearing as follows.
By replacing the masses of $\pi$ in the decay $\sigma\rightarrow\pi\pi$ with the complex mass,
processes like $\sigma\pi^{\rm thermal} \rightarrow\sigma\pi$
are newly taken into account.
Since this process is 
allowed for any energy of the initial $\sigma$ \cite{nishi},
$\pi\pi$ threshold disappears thereby.
Due to the above process, the $\sigma$ peak at the threshold is expected to
acquire a width of 
\bey
\varGamma_{\sigma}\sim n(E_\pi)\cdot\sigma_{\pi\sigma}\cdot m_{\pi}^3
\sim 100\,{\rm MeV}
\label{ordergamma}
\eey
for $T\sim140\,{\rm MeV}$,
where $E_\pi$ is the typical energy of thermal pions.
$\sigma_{\pi\sigma}$ is the scattering cross section of $\sigma$ with $\pi$ and its value 
is estimated to be ${\cal O}(1/f_{\pi}^2)$ from the low energy theorem.
In the present calculation, 
only the mass shift of the internal pion for the one-loop self-energy is taken into account.
One might be worried whether other effects, such as vertex correction and wave function renormalization, might drastically change the results.
The physical point, however, is that the calculated spectral function using the complex pion mass is consistent with the estimate, \eq{ordergamma}.
Therefore, the results of the present paper should be essentially correct even after including other effects.

As was discussed in Ref.\cite{chiku}, the behavior of $\rho_\sigma$ 
is reflected on the diphoton emission rate
from the decay $\sigma \rightarrow 2 \gamma$.
This indicates that the diphoton yield will also be smeared out,
which will be reported elsewhere \cite{ohtani}.
The smeared behavior of $\rho_\sigma$ is also seen for
finite density \cite{chanf,chaos},
though physical origin of the width is different.
It is interesting to apply the prescription adopted here 
in the context of finite density system.

What we have observed in this paper is expected to be universal for processes in which a particle decays into unstable particles.

One of the authors (M.~O.) would like to thank H.~Fujii for fruitful comments and discussions.
The authors would like to acknowledge valuable discussions with T.~Hatsuda.


\def\Ref#1{[\ref{#1}]}
\def\Refs#1#2{[\ref{#1},\ref{#2}]}
\def\npb#1#2#3{{Nucl. Phys.\,}{\bf B{#1}},\,#2\,(#3)}
\def\npa#1#2#3{{Nucl. Phys.\,}{\bf A{#1}},\,#2\,(#3)}
\def\np#1#2#3{{Nucl. Phys.\,}{\bf{#1}},\,#2\,(#3)}
\def\plb#1#2#3{{Phys. Lett.\,}{\bf B{#1}},\,#2\,(#3)}
\def\prl#1#2#3{{Phys. Rev. Lett.\,}{\bf{#1}},\,#2\,(#3)}
\def\prd#1#2#3{{Phys. Rev.\,}{\bf D{#1}},\,#2\,(#3)}
\def\prc#1#2#3{{Phys. Rev.\,}{\bf C{#1}},\,#2\,(#3)}
\def\prb#1#2#3{{Phys. Rev.\,}{\bf B{#1}},\,#2\,(#3)}
\def\pr#1#2#3{{Phys. Rev.\,}{\bf{#1}},\,#2\,(#3)}
\def\ap#1#2#3{{Ann. Phys.\,}{\bf{#1}},\,#2\,(#3)}
\def\prep#1#2#3{{Phys. Reports\,}{\bf{#1}},\,#2\,(#3)}
\def\rmp#1#2#3{{Rev. Mod. Phys.\,}{\bf{#1}},\,#2\,(#3)}
\def\cmp#1#2#3{{Comm. Math. Phys.\,}{\bf{#1}},\,#2\,(#3)}
\def\ptp#1#2#3{{Prog. Theor. Phys.\,}{\bf{#1}},\,#2\,(#3)}
\def\ib#1#2#3{{\it ibid.\,}{\bf{#1}},\,#2\,(#3)}
\def\zsc#1#2#3{{Z. Phys. \,}{\bf C{#1}},\,#2\,(#3)}
\def\zsa#1#2#3{{Z. Phys. \,}{\bf A{#1}},\,#2\,(#3)}
\def\intj#1#2#3{{Int. J. Mod. Phys.\,}{\bf A{#1}},\,#2\,(#3)}
\def\sjnp#1#2#3{{Sov. J. Nucl. Phys.\,}{\bf #1},\,#2\,(#3)}
\def\pan#1#2#3{{Phys. Atom. Nucl.\,}{\bf #1},\,#2\,(#3)}
\def\app#1#2#3{{Acta. Phys. Pol.\,}{\bf #1},\,#2\,(#3)}
\def\jmp#1#2#3{{J. Math. Phys.\,}{\bf {#1}},\,#2\,(#3)}
\def\cp#1#2#3{{Coll. Phen.\,}{\bf {#1}},\,#2\,(#3)}
\def\epjc#1#2#3{{Eur. Phys. J.\,}{\bf C{#1}},\,#2\,(#3)}
\def\mpla#1#2#3{{Mod. Phys. Lett.\,}{\bf A{#1}},\,#2\,(#3)}
\def\etal{{\it et al.}}



\end{document}